\documentstyle[aps,prl,epsf,twocolumn]{revtex}
\begin{document}
\def\bea{\begin{eqnarray}}
\def\eea{\end{eqnarray}}
\def\a{\alpha}
\def\D{\langle l \rangle}
\def\p{\partial} 
\def\break#1{\pagebreak \vspace*{#1}} 
\draft
\title{ Bloch Walls and Macroscopic
String States in Bethe's solution of
the Heisenberg Ferromagnetic  Linear Chain } 
\author{Abhishek Dhar$^{1}$ and B Sriram Shastry$^{2}$ } 
\address{  $^1$ Theoretical Physics Group, Raman Research Institute,
 Bangalore- 560080, India. \\
$^2$ Department of Physics, Indian Institute of Science,
Bangalore 560012, India.\\}
\date{\today}
\maketitle
\widetext
\begin{abstract}
We present a calculation of the lowest excited states of the Heisenberg
ferromagnet in 1-d for any wave vector. These turn out to be 
string solutions of Bethe's equations with a macroscopic number of
particles in them. These are identified as generalized quantum Bloch wall
states, and  a simple physical picture provided for the same.
\end{abstract}


\narrowtext
{ \it Introduction} The question of  elementariness of excitations
in low dimensional magnetic systems is  receiving much attention
currently. It was perhaps first addressed  in the seminal work of
Bethe in 1931\cite{bethe}. In addition to providing the celebrated
 {\it Ansatz} named after him, Bethe asked if Bloch's magnons
are the ``most elementary'' excitations in 1-d. He came to the conclusion 
that they were not, and instead found that the bound states of spin reversals
were. After the original paper of Bethe, the ferromagnet has received
\cite{ckm,bethefinitet,bill}
 comparatively less attention than its antipode, namely the 
antiferromagnet\cite{yang,gaudin}. 
One source of  revival of interest in the ferromagnet
is in connection with stochastic dynamical systems,
albeit with a complex  Aharonov Bohm magnetic flux\cite{dhar,spohn}.
Another  notable recent  exception is a work by Sutherland\cite{bill}, 
who shows that the excited states of the ferromagnet contain 
a singlet state at momentum $\pi$, with an excitation energy
(EE) that is very low, of $O(1/N)$, where $N$ is the length of the ring.

At the semiclassical  level,  domain wall arguments
\cite{blochwall} lead one to expect  in dimensions $d$ ( with volume $=N^d$)
the {\it Bloch wall} excitations to be of $O(N^{d-2})$ , and hence to be amongst
contenders for the lowest  EE in $d=1$. Such  ``large deviation'' excitations
carry spin as well as momentum as we show below. 
These configurations of spins will be 
discussed within the context of Bethe's Ansatz (BA) for the $s=1/2$ 
ferromagnet presently.  

In this work we ask ( and answer) the following question:
{\it For  a given value of the total momentum or total spin of 
the Bethe ferromagnet,  what is the lowest  excited state?} 
  The
nontriviality of the question arises from  the fact that within 
the famous Bethe
formula for the boundstate of $n$ magnons,
 $\omega_{Bethe}(q) = J \frac{2}{n} ( 1- \cos(q) )$,
the lower limit on the total momentum $q$   depends implicitly  upon $n$. Its
dependence  has not been
fully explicated earlier, at least as far as we could find in the literature.
In this work, we use a combination of exact diagonalization and analytic methods to
attack the problem. A new and essential tool that we develop is the argument
of continuity of certain  regular root solutions with 
\break{0.9in}
respect to the density, regarded as a continuous variable,
 leading to a  differential equation
formulation of Bethe's equations (BE) that bypasses the knowledge ( or
otherwise ) of the quantum numbers. 

Our findings are readily stated:  
The lowest excitations for any  momentum $q$ arise
from  special string solutions of BE. These special string solutions
involve {\it a macroscopic number 
of particles} in a given string, and hence we call them macroscopic strings.
Sutherland's solution at momentum $\pi$ is a particular case of these. 
We have found the lowest solution for every value of the total momentum,
these correspond to a definite value of total spin as well.
 Such states
are of the type that one would expect from Bethe's formula for $n$ magnon 
bound states,  with $n$ of the size of the lattice. 
The formula of Bethe cannot, however, be used for such large bound states
since we show that there are significant corrections to the traditional
assumption of a  uniformly spaced  vertical Bethe
string in the complex plane: the curvature
and nonuniformity of spacing produces
essential differences.
Our states can be represented by a new  formula: $\omega_{BW}(q)= \frac{ 2 \pi}{N} J q( 1-
q/(2 \pi) )$ for $ 0 \leq q \leq   \pi$. We find that the solution for a given
$q$  corresponds to a particular value of total spin, 
$S_{tot}= N(1/2 - {q}/{2 \pi})$.
If we write $n$ the spin  deviation from the  saturated ferromagnet ( $S^z=
N/2 -n$) in terms of the density $d$ as  $n= d \ N$, the spectrum can be written
as $\omega_{BW}= \frac{4 \pi^2}{N} J  d ( 1-d)$ with $ q = 2 \pi d$ and $S_{tot}= 
N({1}/{2} -d )$. 
We show finally that these states correspond  to generalized quantum  Bloch walls\cite{comment1}:
{\it thus the  states with lowest EE at any wave vector of $O(1)$ are  Bloch walls}\cite{comment2}.

 Before presenting our calculations, we also note that 
the bound states of Bethe for $s=1/2$  have  been identified recently with 
solitonic excitations\cite{solitons}
of the non linear classical, i.e. large $s$, Landau Lifschitz equations
$\dot{\vec{S}}(x)= \vec{S}(x) \times \partial^2/\partial x^2 \vec{S}(x) $.
Several explicit solutions of these are known \cite{jevicki,fogedby}.
  We note that Bloch walls form a  certain class of 
exact solutions\cite{solitons,fogedby}of the nonlinear
classical equations , namely   non linear spin waves.

\narrowtext
{\it Calculations}: We consider the ferromagnetic Heisenberg Hamiltonian: 
${\mathcal H}= 2 J  \sum_{l=1,N} [s^2 - \vec{S}_l \cdot \vec{S}_{l+1}
]$, where  for most part we consider $s=1/2$ and $\vec{S}_l=
\frac{1}{2} \vec{\sigma}_l$, in terms of  the usual Pauli spin
operators.

Our first result is from observations on wavefunctions of the 
energy eigenstates obtained from exact numerical diagonalization of
chains upto length $N=16$ in a momentum resolved basis.
We observed  that in all cases,
 the lowest state at momentum $q=2 \pi n/N$ has total angular
momentum $S=N/2 - n$. This implies  that this state can be obtained in the $n$
particle sector where it is a ``maximal'', i.e. highest 
weight  state. The corresponding Bethe
wavefunction has all pseudo momenta nonvanishing, and the momentum
and particle density  related as $q=2 \pi d$.

We  next write the BE for the Heisenberg chain in the
Orbach parametrization as:
\bea
N f( \alpha_l ) = 2 \pi I_l + \sum_{m \neq l} f(
(\alpha_l-\alpha_m)/2 )~~~~~~~l=1,2...n 
\label{beteq}
\eea
where $f(x)=\frac{1}{i} \log(\frac{x+i}{x-i}) = 2 \mbox{ArcCot}(x) $, $\alpha_l =
\cot(k_l/2) $ and $\{ I_l \} $ are the Bethe integers, and
$k_l$ the Bethe pseudo momenta. We take the
branch cut of $f(x)$ to be on the imaginary 
axis running from $-i$ to $+i$. 
The energy and total momentum are given respectively by: $
\epsilon= J \sum_{l=1,n} \frac{4}{1+\alpha_l^2}$ and $q=\frac{2 \pi
} {N} \sum_{l=1,n} I_l $.
Let us note that we are interested in the lowest energy states for a given
$q$, requiring a knowledge of the integers $I_l$. These integers, as shown
by Bethe, differ by 2 for scattering states, and by either one or zero for 
bound states in general. Eliminating the integer sets with zeros in them, a very
plausible state is one with $I_l=1 $ for $1 \leq l \leq n$, and indeed we found from
numerical studies of BE
for  small N and small n ( with $n<< N$), that this was indeed so: the resulting state
is invariably the  lowest energy state for small $q$. Emboldened by this
exercise we found the  following exact solution analytically in the 
limit of a thermodynamic
$n$ as well as $N$,  but  at low density  i.e. $d =n/N <<1$ .
Since we are interested in excitation 
energies that are vanishing in the thermodynamic limit, 
the corresponding  variables $\alpha_l$ scale with system size and it is convenient to
introduce new scaled variables $ z_l=\alpha_l/n $. In terms of these,
the lhs of Eq(\ref{beteq}) becomes 
${2}/{(z_l d)}$ on  
using the large $x$ expansion of $f(x)$ and ignoring terms of size
$O(1/N^2)$. 
 On the rhs we cannot make the expansion in general
since, at high densities, a core is formed\cite{bill}  where the separations $n
(z_l - z_m)/2 $ between pairs of particles can become arbitrarily
close to the branch points of $f(x)$.  In  the
low density limit  we find it is possible to
obtain a perturbative solution, using the 
crucial observation  that the  typical interparticle separation 
$ \sim 1/\sqrt{d} $, and hence we can use the smallness of $d$
as an expansion parameter in a perturbative sense. Setting $I_l=1$ we find
the approximate equation:
\bea
\frac{1}{z_l}= \pi d + \frac{2 d}{n} \sum_{m \neq l}
\frac{1}{z_l-z_m}.
\label{ldeq}
\eea    
Note that Eq(\ref{ldeq}) has corrections from the expansion 
of the phase shift, that is typically of $O(d)$ smaller than the least term
retained. 
The solution of the system Eq(\ref{ldeq}) can actually be
found exactly\cite{bss}, but we
save it for a future publication. We find at low densitites
the following result:
\bea
z_l=\frac{1}{\pi d} + \frac{i \sqrt{2}}{\pi \sqrt{d}} x_l-
\frac{2}{3 \pi} (x_l^2+1-\frac{1}{n}) + O(\sqrt{d}),
\label{ldroots} 
\eea
where $x_l$ satisfy $H_n(\sqrt{n} x_l) = 0$, $H_n$ being the $n$th
order Hermite polynomial.
In the limit of large $n$ the $x_j$ form a continuum stretching from
$- \sqrt{2}$ to $\sqrt{2}$ with the familiar semicircular 
density of states $\rho(x)= \frac{1}{\pi} \sqrt{2 - x^2}$.
 This solution can be used
to obtain the energy to order $d^2$. 
The energy $ N \epsilon = 
4J/(n d) \sum_i 1/ z_i^2 $ for low $d$ can be
 found as $N \epsilon = 4 J \pi^2 [ d + d^2 \{ 3 <(\beta_i^{(1)} )^2> -
2 < \beta_i^{(2)} > \} + O( d^3) ]$, where the averages are normalized
sums over the indicated variables.
 Using the explicit expression Eq(\ref{ldroots}) and  converting 
the sums to integrals
 over the  semicircular density of states we get finally the low density formula:
$N \epsilon= 4 J \pi^2 d (1-d) + O(d^3)$. Below we will argue that  
there are no corrections to the above formula beyond the first term: {\it it
is exact!}. Thus provisionally we write
\bea
N \epsilon= 4 J \pi^2 d (1-d)   =2 J \pi q (1-\frac{q}{2 \pi}).
\label{form}
\eea
We note that the low density Equation(\ref{ldeq}) must be abandoned once the
minimum separation $n ( z_i- z_j)$ hits the value $2 i$, this happens at
$2= |n(z_0- z_i)|= \frac{\sqrt{2}}{\sqrt{d} \pi} n|(x_0-x_1)$. However
 $ n|(x_0-x_1)|= 1/\rho(0) = \pi/ \sqrt{2}$, thus $d \sim 1/4$.  Indeed we found
for small systems  that $ d \geq 1/4$ cannot be treated easily numerically:
  the  new difficulty is  that the quantum numbers are no longer simple as
 we discuss below. For $d \leq 1/4$ 
the low density result Eq(\ref{ldroots}) and the full solution of
Eq(\ref{ldeq}) \cite{bss} are extremely close.  
 
We now discuss the techniques used for solving Eq(\ref{beteq})
numerically at larger densities. There are two main problems. One is
that the integers jump around
in a complicated way that is  not known  beforehand
 and it is clearly not feasible to
try all combinations. The second problem is the formation of the core\cite{bill},
i.e. successive  roots that are placed very close to a separation $ 2 i$,
which causes singularities in the equation and results in numerical
inaccuracies. 
Our strategy is to start from low densities where we know the 
roots, and change $d$ slowly and study the evolution of the
roots. 
For this we first convert Eq(\ref{beteq}) into a set of first
order ordinary differential equations (ODE) with $\log(d)$
 as a time like flow-parameter. Taking the derivative of
Eq(\ref{beteq}) with respect to $d$ and defining new 
variables $t= \log{d}$ and $f_l=z_l e^{t}$ we obtain: 
\newpage
\bea
\sum_m &A_{lm}&\dot{f_m} = -\pi f_l^2 \dot{I_l}-\sum_{m}
A_{lm} (f_l-f_m),
\eea
\bea
{\rm{where}}~&A_{lm}&=\frac{2 e^{t} f_l^2}{n} \frac{1}{[4 e^{2 t}/n^2 
+ (f_l-f_m)^2]},~ m \neq l \nonumber \\
&A_{ll}&=1-\sum_{m \neq l} A_{lm} \nonumber
\eea
and the derivatives are with respect to the ``time''  variable $t$.
The time derivatives of the integers are delta functions and hence
drop out of the equations at almost all times. Also since the flow of
the roots themselves is smooth, we expect the delta-function
singularities to be {\it precisely cancelled} by other terms in the
equation. Hence in evolving the above ODE we can drop the first term
on the right hand side of the equation. This immediately solves the
problem of our lack of knowledge of the integers since they do not occur
anywhere else in the differential equations.
From these solutions we can recover the integers and use
them in the root-finder to get more accurate
solutions. We find that till densities around $d=0.45$ the solutions obtained
from the ODE are very accurate. At higher densities,
significant numerical errors show up because of the singularities
associated with the core. In these cases we correct our ODE solutions
by fixing the core by hand and using the root-finder to self-consistently
solve for the roots outside the core. We plot in Fig. \ref{solu} the
solutions, for a system of $16$ particles, at different densities.   
The table below shows the integer sets at four different densities
(since they occur in pairs we show only half of them):

\vspace{0.25cm}
\begin{tabular}{|c|c|c|c|c|c|c|c|c|}  
\hline 
~~d=0.2~~ &~~1~  &~~1~ &~~1~ &~~1~ &~~1~ &~~1~ &~~1~ &~~1~  \\  \hline 
~~d=0.3~~ &~~1~  &~~1~ &~~1~ &~~0~ &~~1~ &~~1~ &~~2~ &~~1~  \\ \hline 
~~d=0.4~~ &~~1~  &~~0~ &~~1~ &~~0~ &~~1~ &~~1~ &~~2~ &~~2~  \\ \hline 
~~d=0.5~~ &~~0~  &~~0~ &~~0~ &~~0~ &~~0~ &~~0~ &~~0~ &~~8~  \\ \hline 
\end{tabular}
\vspace{0.25cm} 

We now discuss the energies that we obtain from the BA solutions. 
For $N \le 16$, we have verified that all the solutions
obtained from the numerical solutions of the BE using the
above scheme, match with those obtained from exact numerical
diagonalization. With the BE we can go to much larger
system sizes. {\it We find that the gap vanishes at every finite $q$,
with system size dependence $\sim 1/N$.} In Fig. \ref{disp} we plot the
system size dependence of the 
gap at two densities, namely at quarter and half fillings. The latter
case corresponds to the $q=\pi$ state considered by Sutherland and we
verify his result $N \delta E= J \pi^2 $. At $d=1/4$ $N
\delta E$
seems to asymptote to the value $3 J \pi^2/4$. Remarkably both these
asymptotic values of $N \delta E$ at densities $1/2$ and $1/4$ can be
obtained from the low density formula in Eq(\ref{form}).
In Fig. \ref{disp} we also plot the energy-wavevector curve, obtained
from the solution of the $16$-particle problem and compare it with
Eq(\ref{form}). Note that the discrepancies  at large $q$ are finite
size effects and would vanish in the $N \to \infty $ limit. 

{\it Variational results}
Having found the excited states, 
we now  turn to the explicit connection with Bloch wall states. 
We now show, remarkably enough,  that the expression
Eq(\ref{form}) for the gap can 
 be obtained from a  simple variational calculation.
We work with
arbitrary spin $s$  of the particles.
A  neat way to
generate Bloch walls is via a unitary  rotation operator\cite{lsm} 
$ Q= \exp {i \frac{2 \pi}{N} \sum_m m S^z_m} $ acting upon
an appropriate state,
$|0_n> \equiv (\hat{S}^-_{0})^n |ferro>$, 
where $\hat{S}^-_{q} \equiv \sum_j \exp( i q.r_j) S^-_{j}$ is a spin wave 
creation operator carrying momentum $q$ and $|ferro>$ is the 
state with all spins up.
Using  $[H, \hat{S}^-_{0}]=0$,
we see that $|0_n>$ is  the ground state in the
$n$-particle sector with zero total momentum.
Thus finally we  write the variational Bloch Wall state $|BW> \equiv Q |0_n>$.
Using  the quasi commutator $Q \hat{S}^-_{q} = \hat{S}^-_{q + 2 \pi/N} Q$,
we find
that $|BW> = (\hat{S}^-_{ 2 \pi/N})^n |ferro> $,
and  has total momentum $q=2 \pi d$.
In the semiclassical limit, $s >>1$, the above state is readily visualized
as classical spins that are tipped from the z axis and rotate along a cone.
 The variational calculation
of the excitation energy $\delta E$, i.e. 
 $<0_n| Q^{\dagger} {\mathcal H} Q |0_n>$ can be done easily by transforming
the rotation onto the spin operators. Writing $n= d N$, with $0 \leq d \leq 2s$,
 we find $\delta E = \frac{4 J \pi^2}{N} d( 2s -d)$. At $s=1/2$ {\it this is
 also the result of the calculation in}  Eq(\ref{form}). 

At this point we admit that we were  surprised, as the reader might well be,
that the results of an elaborate bound state calculation with 
a macroscopic number of complex roots agrees with the result of a simple
looking variational 
wavefunction that resembles a Bose condensate of spin waves.
This phenomenon  is presumably a consequence of 
the shallow nature of the bound state. We note that the
variational states satisfy
$\sqrt{<{\mathcal{H}}^2>-<{\mathcal{H}}>^2}/<{\mathcal{H}}> \sim
O(1/\sqrt{N})$ which shows that in the $N \to \infty$ limit these
become exact eigenstates, thereby providing independent
evidence for the exactness of the main result of our work.

For large $s$, the energy as well as momentum of these
states agrees with the semiclassical estimates\cite{fogedby} using a Poisson bracket
structure to construct a semiclassical momentum operator. 
The variational results for all values of $s$
 thus collapse onto the same formula. The conjecture of \cite{jevicki,fogedby}
implies just this kind of a  result, but for the solitons, i.e. for bound states
with small number of spin deviations. Needless to say, we believe that our
variational results are exact (in the thermodynamic limit) for {\it all spin},
since we have established  them at $s=1/2$ and also for very large $s$.

We note that the Bloch wall states $|BW>$ carry a total spin that is easy to 
calculate using simple extension of the above calculation: 
 $<S^2_{tot}> = <S_z>^2=N^2 (s-d)^2$.
Finally we note that these variational results for Bloch walls are 
generalizable to higher dimensions.
 
{\it Concluding remarks} 
We  finally note in summary that the Bethe formula for $n$ magnon bound states:
$\omega_{Bethe}= J\;  \frac{2}{n} ( 1- \cos(q))$
 is (a) exact for $q >> 2 \pi n /N$, (b) invalid for $q < 2 \pi n /N$
and (c) has significant corrections when 
 $ q \sim 2 \pi n /N$ \cite{ftnt}. The evidence for (a)
is in Bethe's paper itself, the corrections to it are exponential in N,  (b)
is numerical.  Evidence for (c) has been presented in this paper, where we
find instead: $\omega_{BW}= 2 \pi J/N  |q|( 1- \frac{|q|}{ 2\pi})$.

$^1$ Also at the Poornaprajna Institute, Bangalore. dabhi@rri.ernet.in 
$^2$ Also at the JNCASR, Bangalore. bss@physics.iisc.ernet.in

\vbox{
\vspace{0.5cm}
\epsfxsize=8.0cm
\epsfysize=6.0cm
\epsffile{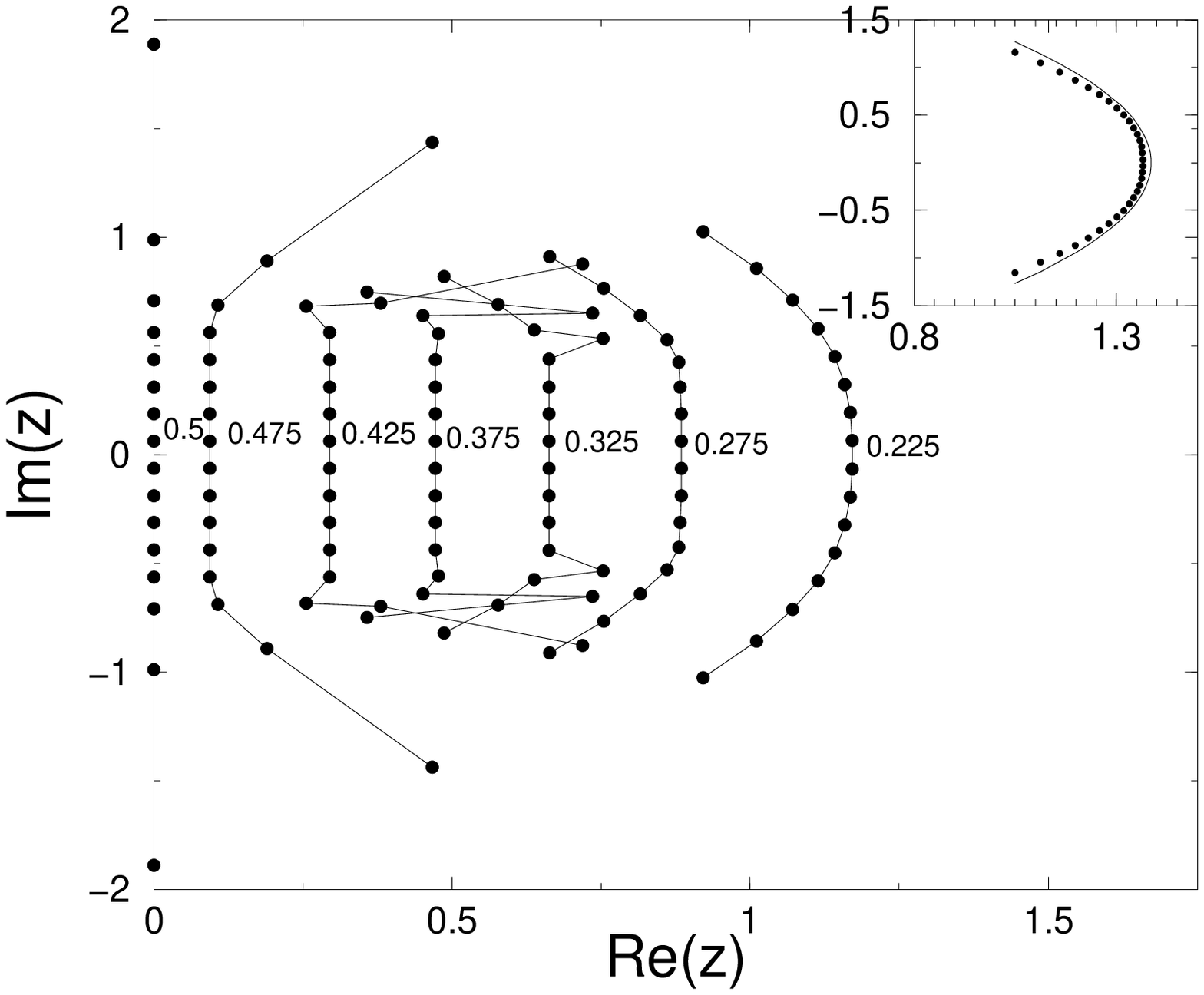}
\begin{figure}
\caption{\label{solu} The figure shows the Bethe curves for a
$16$-particle system at different densities (values indicated along
each curve). The inset compares the numerically 
obtained roots (points), for a $32$-particle system at $d=0.2$, with the
exact low-density result (solid line).    
}  
\end{figure}}
\vbox{
\vspace{0.5cm}
\epsfxsize=8.0cm
\epsfysize=6.0cm
\epsffile{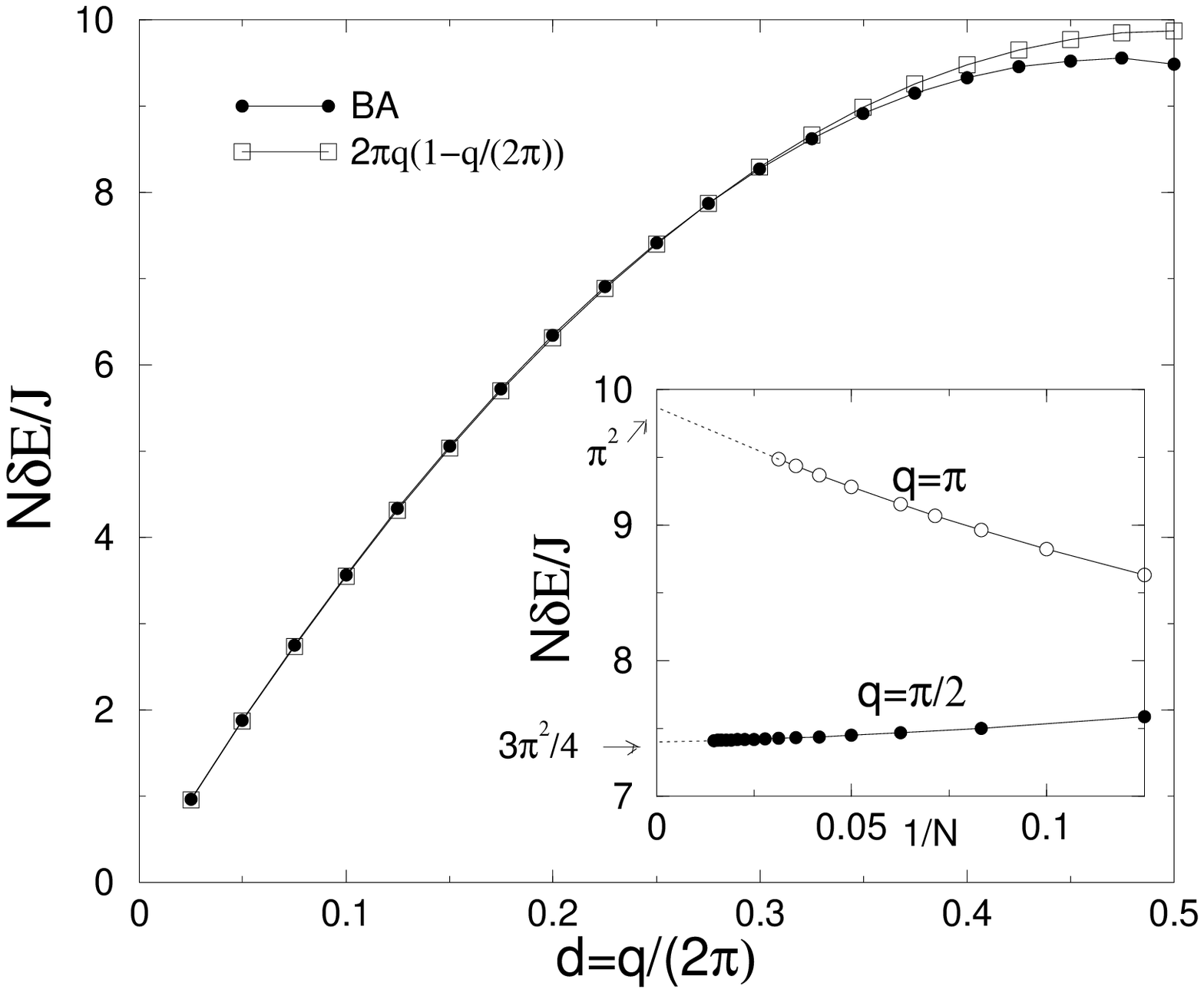}
\begin{figure}
\caption{\label{disp}  The gaps at different wave vectors as obtained
from the solution of the BA equations for $16$ particles.
The gap as given by Eq(\ref{form}) is also shown.
The inset shows dependence of the gap on system size at
two different wave vectors.
}  
\end{figure}}


\begin{references}
\bibitem{comment1} We use the term Blochwall in  a slightly generalized sense from the
usual domain walls of \cite{blochwall}. In the latter, the magnetization rotates from
up at $x=0$ to down at $x=L$, thus defining  spin-twisted  boundary conditions.
 Our excitations satisfy periodic boundary conditions, with the magnetization
rotating along   a cone as we traverse the ring, and hence constitute $2,4,6,\ldots$ domain walls of the former.
\bibitem{comment2} For wave vectors $q$ of $O(1/N)$, spin wave energies are
on a scale $O(1/N^2)$, 
and these are also obtained from our fromula Eq(\ref{form}). We mention that
the very lowest excited state ( over all $q$)
 is the Bloch spin wave with $q= \frac{2 \pi}{N}$, with the usual energy $ J 4 \pi^2/N^2$,
much above the lower bound $4J/N^2$ found in \cite{spohn}.
\bibitem{diffeq} This procedure of using the density of particles as a
continuous variable and writing a differential equation for the evolution of
the roots, has, to our knowledge, not been used earlier. It is crucial in
circumventing the problem of lack of knowledge of the Bethe integers.
\bibitem{ftnt} Note that this is true even for small $n$. For
example for $n=2$, $q=4 \pi/N$ the roots, from Eq(\ref{ldroots}), are at
$\alpha_{1,2}=N/{\pi} \pm i \sqrt{N}/{\pi}$  instead of Bethe's $N/\pi \pm
i$. As noted by Bethe, for the 2-magnon case the string solution
holds for $q \stackrel{>}{\sim} 1/\sqrt{N} $.
\end{references}
\end{document}